# Deneb and the α Cygni Variables

*Joyce A. Guzik*
*Los Alamos National Laboratory*
*MS T-082, Los Alamos, NM 87545, USA*
*joy@lanl.gov*

*Brian Kloppenborg*
*American Association of Variable Star Observers*
*185 Alewife Brook Parkway, Suite 410*
*Cambridge, MA 02138, USA*

*Jason Jackiewicz*
*Department of Astronomy, New Mexico State University*
*PO BOX 30001, Las Cruces, NM 88003, USA*

**Abstract**

Deneb, the prototype α Cygni variable, is a blue-white supergiant that shows irregular variability with quasi-period around 12 days in brightness and radial velocity. Abt et al. (2023) found that larger amplitude 12-day variations appear to resume abruptly and at an arbitrary phase and damp out after several cycles, with an interval of around 70 days between these resumptions. Here we make use of an 8.6-year photometric data set for Deneb from the Solar Mass Ejection Imager (SMEI) to better characterize this behavior. We find that the interval between pulsation resumptions is not exact, with the most common intervals between 100 and 120 days. Sometimes one or more intervals are skipped. We also examine AAVSO and Transiting Exoplanet Survey Satellite (TESS) light curves for α Cyg variables Rigel, Saiph, and Alnilam in Orion, Aludra in Canis Major, and 6 Cas to compare with the behavior of α Cyg. Except for 6 Cas, the time series are too short, or the observations too infrequent to draw any conclusions about similarities between the behavior of these stars and α Cyg. We also summarize results of evolution and pulsation modeling for Deneb and α Cyg variables from the literature. The α Cyg variables may not be a homogenous group with a common mechanism for their variability. It has not been determined whether they are on the first crossing of the Hertzsprung-Russell diagram toward the red supergiant phase or are on their second crossing after having been red supergiants. Future plans include examining BRITE Constellation data for Deneb, processing SMEI data for other bright α Cyg variables, and comparing 6 Cas light curves from AAVSO and TESS data taken concurrently.

## 1. Introduction

Deneb, the prototype α Cygni variable, is a bright spectral-type A2 Ia supergiant that shows a 12-day quasi-period with varying amplitude. Analysis of radial velocity and photometry from 1930 to the present shows evidence of abrupt resumptions of higher-amplitude pulsations occurring with an interval around 70 days (Abt et al. 2023, Guzik et al. 2023). Examples of this behavior are shown in the radial velocity data of Paddock (Fig. 1) and the Transiting Exoplanet Survey Satellite (TESS, Ricker et al. 2015) photometry (Fig. 2).

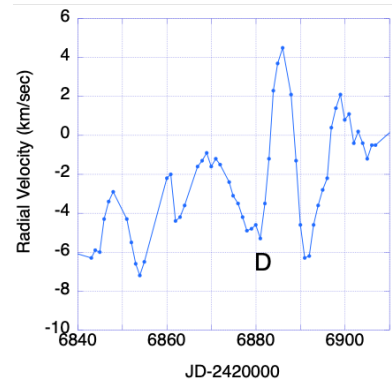

**Figure 1:** Deneb radial velocity data taken by Paddock May 12-July 21, 1932. The point labeled D shows an instance of an abrupt resumption of larger-amplitude pulsations.

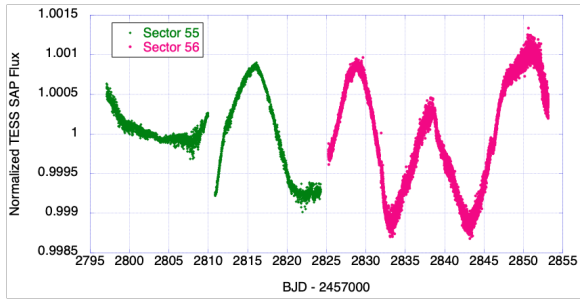

Figure 2: Deneb photometry from the TESS spacecraft August 5-September 1, 2022. Pulsations resume abruptly during Sector 55 (green).

These analyses raise several questions:

- What is the cause of Deneb's pulsations with quasi-period 12 days? Why 12 days?
- Why do the larger amplitude pulsations damp out and resume, perhaps every 70-80 days?
- Do other α Cygni variables show similar behavior?

To begin to address these questions, we compare/contrast the behavior of Deneb with several other bright α Cygni variables. More than one hundred α Cygni variables are listed in the American Association of Variable Star Observers (AAVSO) Variable Star Index (VSX, Watson et al. 2006) database (https://www.aavso.org/vsx/). We make use of an 8.6-year photometric data set from the Solar Mass Ejection Imager (SMEI) to quantify evidence for any 70-day periodic pulsation resumptions in Deneb. We also examine light curves from TESS and AAVSO data for α Cygni variables Rigel, Saiph, and Alnilam in Orion, Aludra in Canis Major, and 6 Cas. We review evolution and pulsation modeling results for Deneb and α Cygni variables and identify simulations that could be carried out using, for example, the open-source MESA code.

## 2. Deneb SMEI Data

After results on Deneb were presented at the 112[th] AAVSO Annual meeting, Kloppenborg offered unpublished photometric data on Deneb from the Solar Mass Ejection Imager taken from 2003 through 2011 (Jackson et al. 2004, Clover et al. 2011). The SMEI spacecraft was in a Sun-synchronous polar orbit with a 102-minute period. While the spacecraft was designed to measure solar coronal mass ejections (CMEs), it also observed almost every bright (V < 6) star. It had three cameras equipped with non-linear optics which provided each camera with a 3 x 120 degree field-of-view of the sky. These data require processing to remove camera-to-camera zero-point offsets, angle-dependent flux loss causing a ~100 day curvature or change in slope, and other spacecraft-related noise and artifacts.

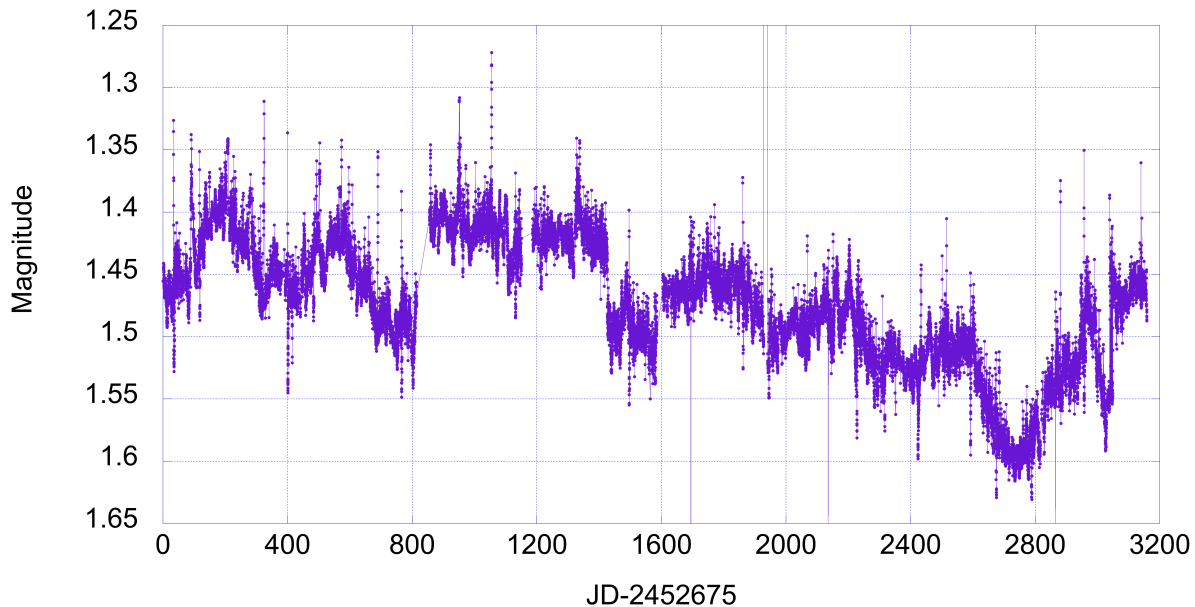

Figure 3: Deneb 8.6 year partially processed light curve from Solar Mass Ejection Imager (SMEI)

## 2.1 Deneb SMEI light curve

Figure 3 shows the 8.6-year Deneb SMEI light curve after some processing to partially remove curvature and offsets. One data point is taken each spacecraft orbit, i.e., every 102 minutes or 1.7 hours. Figures 4-6 zoom in on portions of the light curve where the ~12-day Deneb pulsation period can be discerned. These light curve segments also show some discontinuities of unknown origin that may be intrinsic to the star or may be remnants of incomplete removal of offsets. These discontinuities are outlined by many SMEI data points, and therefore not the consequence of a single bad data point. Since other Deneb time-series data sets also show discontinuities (see, e.g., the Richardson et al. (2011) data plotted in Abt et al. 2023), we are reluctant at this point to rule out intrinsic stellar behavior as the source.

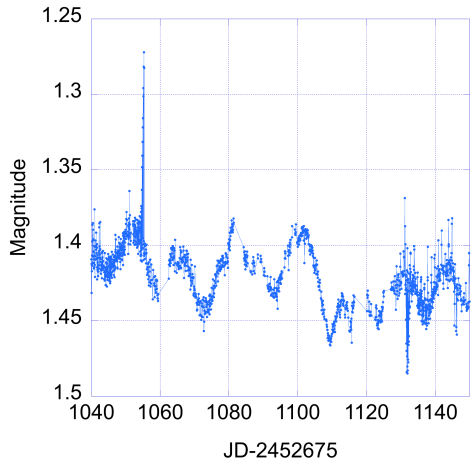

**Figure 4:** Magnitude vs. Julian date for Deneb SMEI data, showing some oscillations with quasi-period 12 days, and two discontinuities

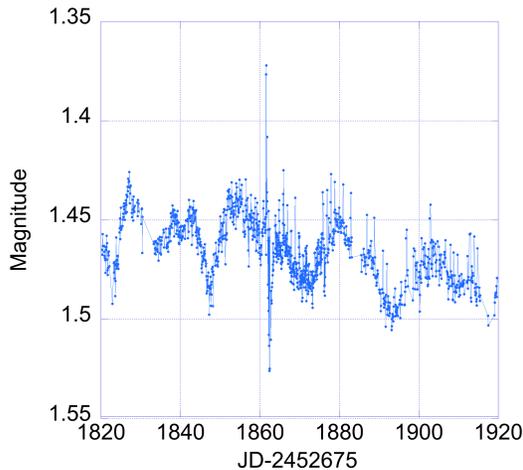

**Figure 5:** Magnitude vs. Julian date for Deneb SMEI data, showing a few oscillations with quasi-period 12 days, and one abrupt discontinuity that seems to interrupt the phase of one of the oscillations

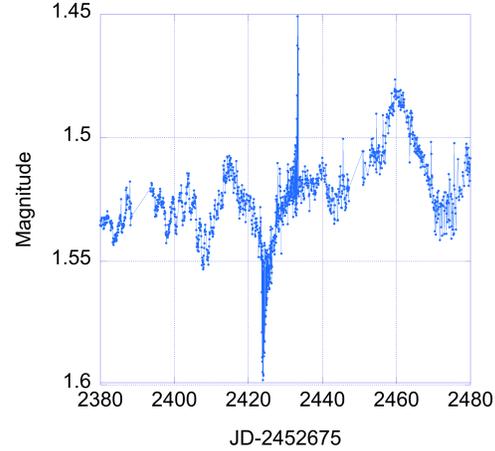

**Figure 6:** Magnitude vs. Julian date for Deneb SMEI data, showing a few oscillations with quasi-period 12 days, and two closely spaced discontinuities

## 2.2 Statistics of events

We attempted to quantify the intervals between resumption of a few cycles of larger-amplitude 12-day pulsations as seen in Figures 4-6, as well as the intervals between the discontinuities. Figure 7 shows a histogram of the days between the discontinuities. The most common interval is in the third bin after 50 days, i.e., between 75 and 87.5 days. This period is close to but a little longer than the intervals suggested by Abt et al. (2023) for abrupt resumptions in pulsations.

Figure 8 shows a histogram of the number of days between resumptions of larger amplitude pulsations with quasi-period 12 days. The interval is less pronounced but appears to peak between 100 and 125 days.

As hoped, the Deneb SMEI data has high-enough cadence and accuracy to identify resumptions of ~12 day pulsations. In the SMEI data, the abrupt excursions, if intrinsic to the star, are not correlated with the resumption of larger-amplitude pulsations. The pulsation resumptions seem to occur at longer and more irregular interval than the approximate 70-day intervals hypothesized by Abt et al. (2023). The abrupt excursions do not appear to 'reset' the pulsation phase, except perhaps in the one example of Figure 5.

We also applied Fourier analysis to the entire SMEI data set. Not unexpectedly, the amplitude spectrum (Fig. 9) does not show any well-defined periodicity.

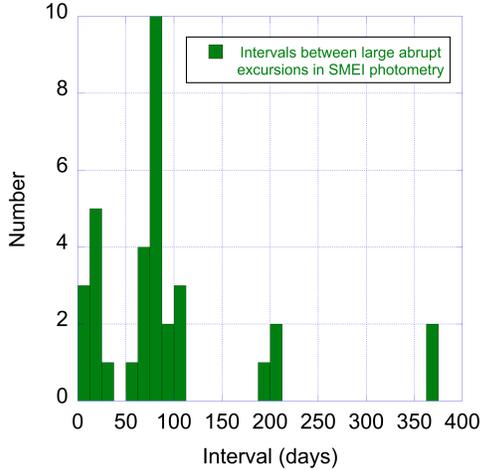

**Figure 7:** Intervals between discontinuities in Deneb SMEI photometry

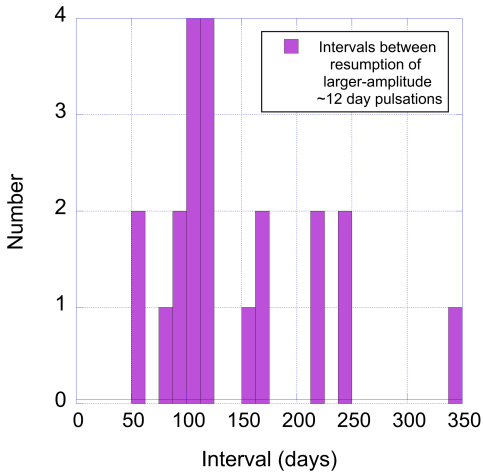

**Figure 8:** Intervals between resumption of a few cycles of ~12-day larger amplitude pulsations in Deneb SMEi photometry

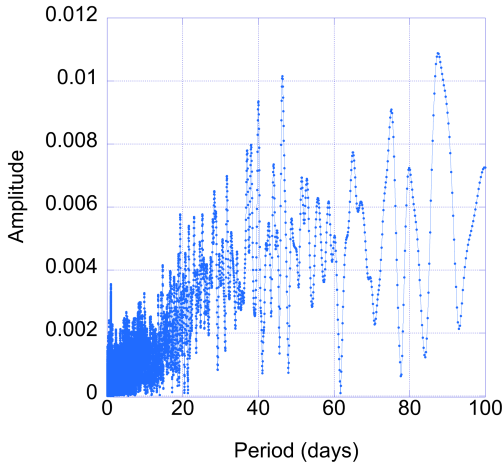

**Figure 9:** Amplitude spectrum of Deneb SMEI data, showing that there is no well-defined periodicity

## 3. Other α Cyg Variables

To begin to answer the question of whether other α Cyg variables have light-curve properties similar to Deneb's, we chose a few of the brightest α Cyg variables from the AAVSO VSX database: Rigel, Saiph, and Alnilam in Orion, and Aludra in Canis Major. We also decided to look at 6 Cas because this α Cyg variable was mentioned in Abt et al. (2023) as having a similar spectral type and age as Deneb's.

We found data for all of these stars in the AAVSO International Database (Kloppenborg 2023). The data for Aludra were sparse, and the data for Rigel, Saiph, and Alnilam did not have the precision to follow pulsational variations in detail. The data for 6 Cas were more extensive and will be discussed below.

We also found TESS data for all of these stars at the Mikulski Archive for Space Telescopes (mast.stsci.edu). We discuss the High-Level Science Product (HLSP) light curves for each star below. These light curves have been processed by various investigators using special pipelines starting from the TESS full-frame images or target pixel files.

### 3.1 Rigel

Rigel (β Ori, HD 34085, TIC 231308237) has V magnitude 0.13 and spectral type B8 Iae. TESS observed Rigel during two 27-day sectors, Sector 5 and Sector 32. However, an HLSP light-curve product is only available for Sector 5. The 30-min cadence light curve (Fig. 10) shows some interesting variations, but a longer time series is needed to identify any quasi-periodicities or to draw any comparisons with Deneb or other α Cyg variables. There are some artifacts remaining in the light curve after data processing (vertical lines at nearly regular intervals).

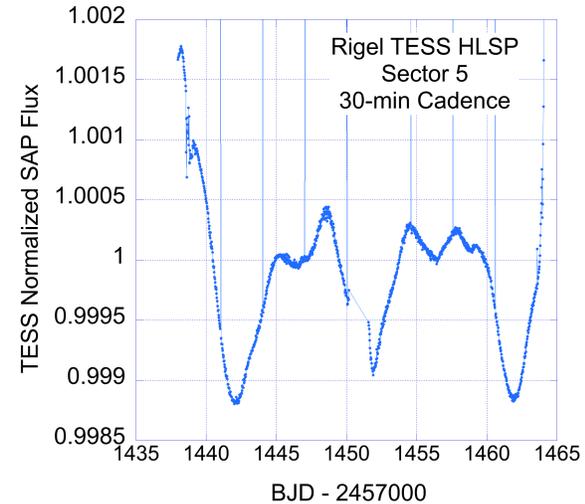

**Figure 10:** Rigel TESS HLSP Sector 5 light curve

## 3.2 Saiph

Saiph (κ Ori, HD 38771, TIC 427451176) has V magnitude 2.06 and spectral type B0.5 Ia. TESS observed Saiph during two 27-day sectors, Sectors 6 and 33. Figure 11 shows the TESS HLSP light curve for Sector 6 with 30-min cadence, and Figure 12 shows the TESS HLSP Sector 33 light curve with 10-min cadence. The Sector 6 light curve has been detrended more successfully than the Sector 33 light curve. Both light curves show aperiodic variability, with some interesting larger excursions to high and low amplitudes. Again, longer time-series data are needed to make any comparisons between Saiph and Deneb.

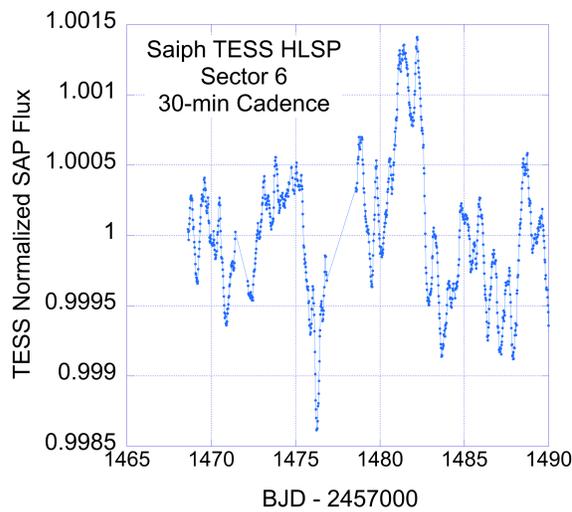

Figure 11: Saiph TESS HLSP Sector 6 light curve

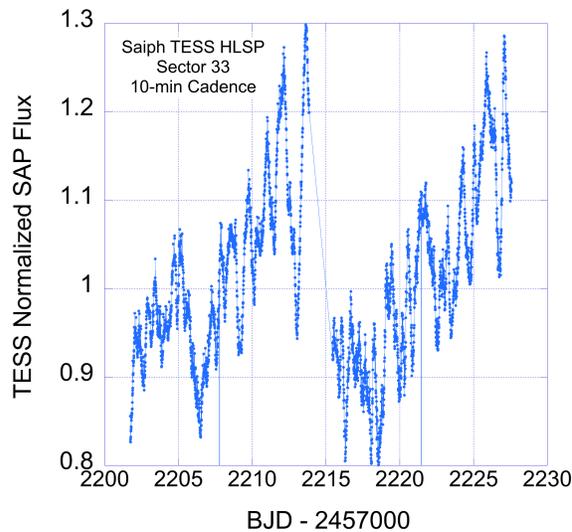

Figure 12. Saiph TESS HLSP Sector 33 light curve

## 3.3 Alnilam

Alnilam (ε Ori, HD 37128, TIC 427451176) has V magnitude 1.69 and spectral type B0 Ia. Alnilam was observed by TESS during Sector 6 and Sector 32. Figures 13 and 14 show the TESS HLSP light curves. A few data artifacts remain, but the HLSP light curves were detrended well. There appear to be some larger dips and peaks in the light curves. A longer time series may help to identify any regularities in the occurrence of these features.

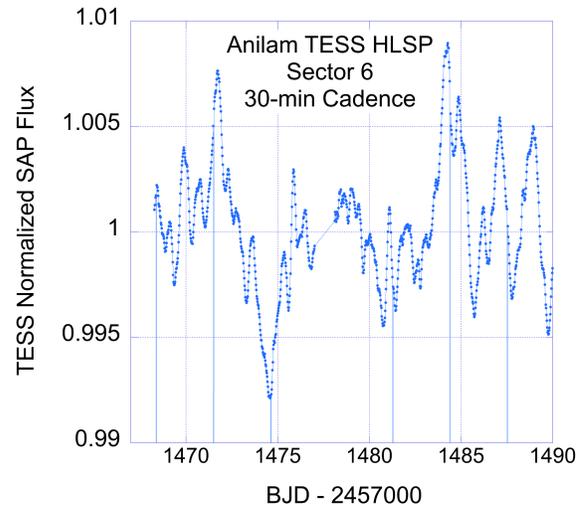

Figure 13: Alnilam TESS HLSP Sector 6 light curve

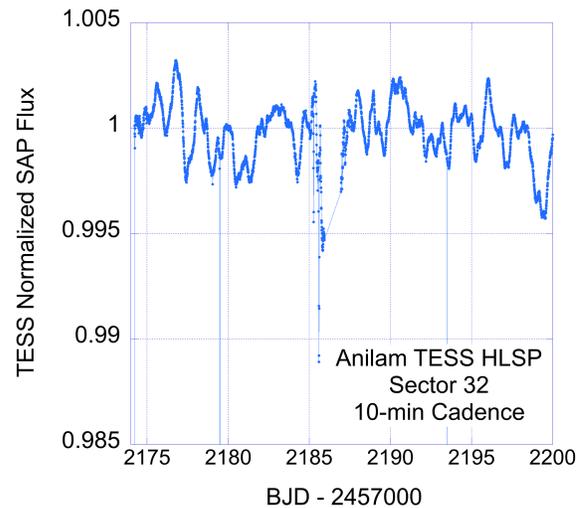

Figure 14: Alnilam TESS HLSP Sector 32 light curve

### 3.4 Aludra

Aludra (η CMa, HD 58350, TIC 107415639) is the bright star in the tail of Canis Major. Aludra has V magnitude 2.45 and spectra type B5 Ia. Aludra was observed by TESS during Sectors 34 and 61. The HLSP TESS light curves (Figs. 15 and 16) appear to have been detrended well, but still contain a few artifacts. As was the case with Alnilam, a couple of high and low brightness excursions are identifiable, but a longer time series is needed to identify any periodicities in the appearance of these excursions.

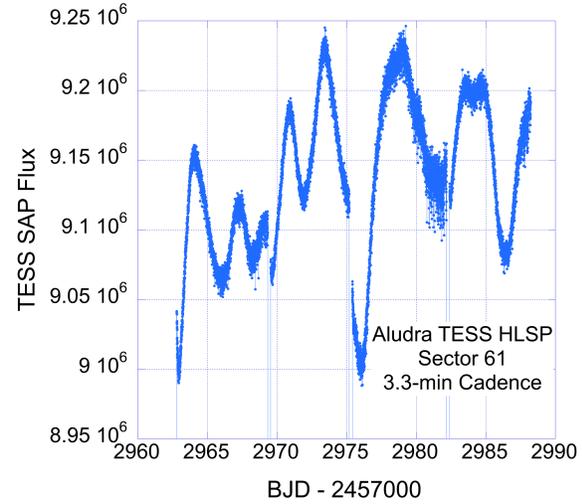

**Figure 16:** Aludra TESS HLSP Sector 34 light curve

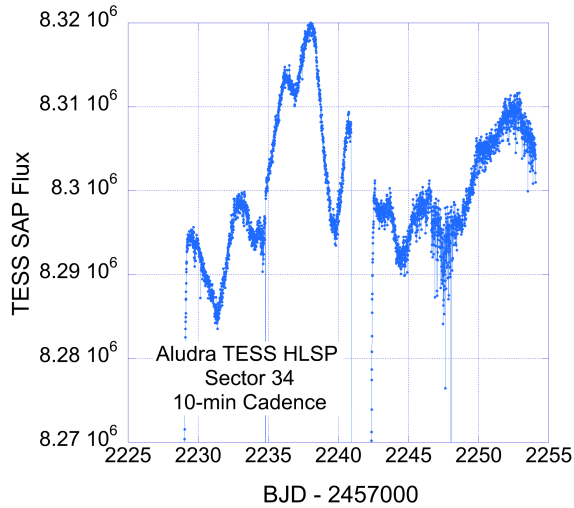

**Figure 15:** Aludra TESS HLSP Sector 34 light curve

### 3.5 6 Cas

6 Cas (V566 Cas, HD 233835, TIC 470694175) is listed as a double or multiple star in the SIMBAD database, with the two stars having spectral types A2.5 I+O9.8 II. Presumably the A2.5 I star, with nearly the same spectral type as Deneb, is the α Cyg variable. The V magnitude of 6 Cas varies between 5.34 and 5.45 with a strongest periodicity of 37 days (Koen and Eyer 2002).

6 Cas has been observed extensively by several AAVSO observers at U, V, B, R, and I wavelengths during the past two years. Figure 17 shows the AAVSO light curves. The ~37-day period is prominent in the observations taken between JD 2459850 and JD 2459910.

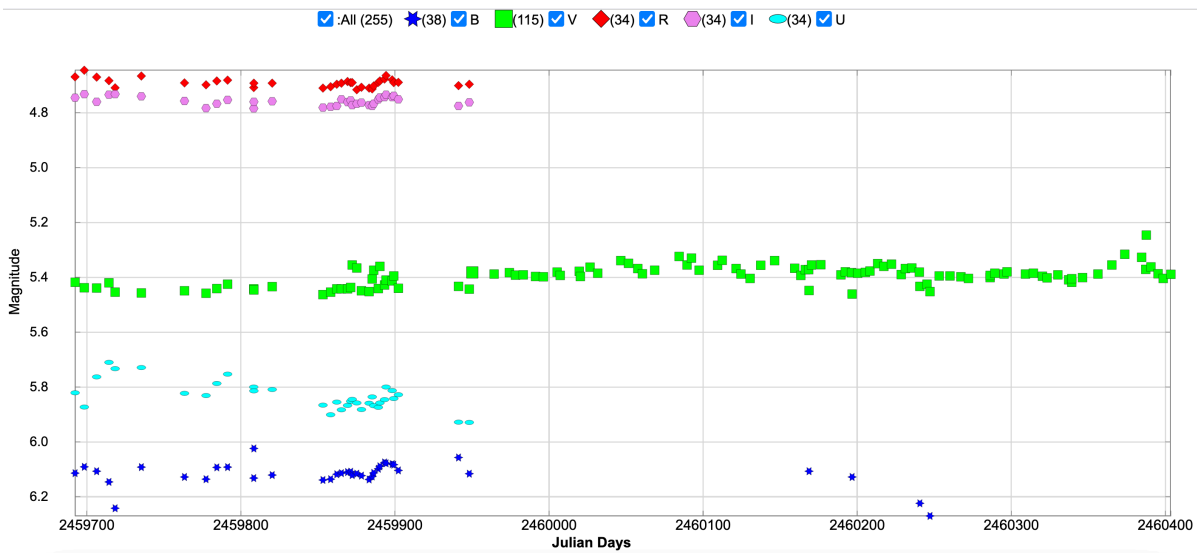

**Figure 17:** 6 Cas light curves from AAVSO database from April 2022 through April 2024

6 Cas was observed by TESS during five sectors: 17, 18, 24, 57, and 58. The HLSP data from these sectors needs further detrending. Figure 18 shows the Sector 18 light curve. Since the time series of TESS sectors is only 27 days, prospects may be better using AAVSO data to look for instances of the 37-day periodicity.

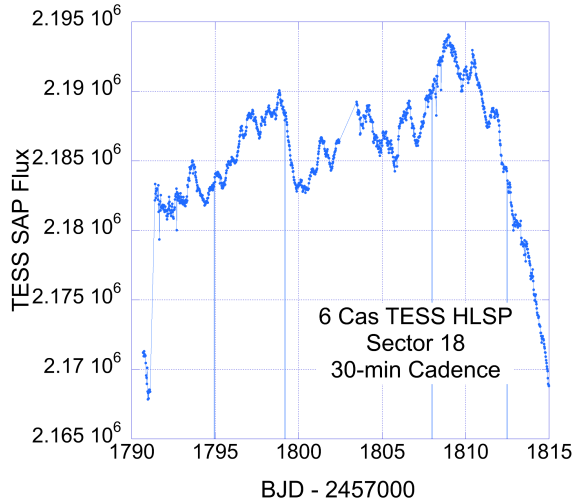

**Figure 18:** 6 Cas TESS HLSP Sector 18 light curve

## 4. Stellar Evolution and Pulsation Modeling

The evolutionary state and cause for variability of α Cyg stars continues to be an area of active research. Some modeling results are found in Gautschy (2009), Georgy et al. (2021), Saio (2011), Saio et al. (2013), and Saio et al. (2016). Evolution modeling of massive stars incorporates treatments for physical processes that are not well constrained, among them, mass-loss rates, rotational mixing, and the treatment of convection. Enhancing mass loss, and to some extent, rotational mixing, can cause evolution models for stars of initial mass greater than 14 solar masses to 'blue loop' back into the blue and yellow supergiant regions of the Hertzsprung-Russell (H-R) diagram after having first evolved across the H-R diagram to become red supergiants.

To explore whether the variability of α Cyg stars is caused by pulsational instabilities, analyses have been carried out to search for mechanisms to excite pulsations with the observed periods of α Cyg variables. Several mechanisms have been proposed and debated. Gautschy (2009) found, for models crossing the H-R diagram for the first time, that nonradial dipole and quadrupole gravity modes are overstable in the stellar envelope above a convection zone overlying the hydrogen-burning shell. Georgy et al., Saio (2011), and Saio et al. (2013, 2016) find that the pulsation properties of α Cyg variables can only be explained by models that have 'blue-looped' after the red supergiant phase. Their analyses find both radial and nonradial modes excited by the kappa (opacity valving) mechanism in ionization regions associated with the Fe opacity bump, $2^{nd}$ helium ionization, or hydrogen ionization, depending on the effective temperature and evolution state. They also find modes excited by so-called strange-mode instabilities for stars with luminosity-to-mass ratio in excess of 10,000 times that of the Sun, and 'low-degree oscillatory convection (nonadiabatic g-) modes' associated with a convection zone produced around 200,000 K by Fe ionization. However, these $2^{nd}$-crossing models have an observational issue--in order to get the models to blue loop, enhanced mass loss, aided by rotational mixing, is needed. This mass loss and mixing, in conjunction with convective 'dredge-up', results in increased surface helium abundances and nitrogen-to-oxygen (N/O) and nitrogen-to-carbon (C/O) ratios at the stellar surface, since helium and nitrogen produced from deep nuclear burning are mixed to nearer the surface and exposed by mass loss. The helium abundance and N/O, and C/O ratios predicted are too high compared to those observed in α Cyg variables. Using the Ledoux criterion for convective stability in the core, instead of the usual Schwarzchild criterion, reduces the amount of convective mixing. This model change almost works to restore agreement with observations for the bluest supergiants, but not for α Cyg.

These authors note that the α Cyg variables may not be an homogeneous class in terms of evolutionary state, and their pulsations may not have a single explanation. The variations may not even be caused by pulsations, but rather by some other surface-layer instability.

The evolution and pulsation results discussed here were conducted with non-open-source codes, but models could be explored (and prior results verified) using the open-source Modules for Experiments in Stellar Astrophysics (MESA, Jermyn et al. 2023) code, (https://docs.mesastar.org/en/release-r24.03.1/), and the associated stellar pulsation codes GYRE (Townsend and Teitler 2013) and RSP (Paxton et al. 2019).

It is likely that modeling codes with capabilities beyond MESA are needed to make progress to understand massive star variability, for example, 2-D or 3-D codes that can model the hydrodynamics of stellar envelopes, or codes that can look at the response of the atmosphere in the presence of strong stellar winds, pulsations, and/or shocks.

## 5. Conclusions and Future Work

The long time-series high-cadence SMEI data was very helpful in answering some of our outstanding questions about Deneb. The larger amplitude ~12 day pulsations do not appear to resume at exact intervals, but sometimes skip intervals, with typical intervals closer to 100 days, longer than the 70-day interval hypothesized by Abt et al. (2023). There does not seem to be strong evidence for abrupt events that interrupt the pulsation phase. There are some discontinuities in the SMEI data which involve multiple data points per event—it is not clear whether these are instrument or data-reduction artifacts, or intrinsic to Deneb's behavior. These events do not appear correlated with the resumptions, and the interval between them is on average shorter, around 75-90 days.

We have examined TESS and AAVSO data for several other α Cyg variables. The TESS data is of high cadence and has excellent precision needed to follow α Cyg variations. The data need further detrending in some cases; the time series lengths are too short to make additional comparisons with the behavior of Deneb.

The AAVSO data for 6 Cas appear promising for investigating properties of its long-term variations. There is even some overlap in time of the AAVSO data with the TESS Sector 57 and 58 data that will be valuable for intercomparison. The other α Cyg variables Rigel, Saiph, Aludra, and Anilam discussed here have some AAVSO data, but the precision is not adequate to do much more than verify that their variations are consistent with α Cyg variables.

We have recently found Deneb data taken using the BRITE constellation satellites (Weiss et al. 2014). We hope to work with the BRITE principal investigators to compare this data to the SMEI data set to confirm some of our conclusions.

Since the Solar Mass Ejection Imager observed many stars brighter than 6$^{th}$ magnitude, we hope to mine this data set for light curves of additional α Cyg variables.

Finally, we may delve into using the MESA code to continue to gain insight on the evolution state and causes of variations for α Cyg variables.

## 6. Acknowledgements

This research made use of data from the Mikulski Archive for Space Telescopes (MAST), the SIMBAD database (https://simbad.u-strasbg.fr/simbad/), and TOPCAT (https://www.star.bris.ac.uk/~mbt/topcat/) software. We acknowledge with thanks the variable star observations from the AAVSO International Database contributed by observers worldwide and used in this research. J.G. acknowledges support from Los Alamos National Laboratory, managed by Triad National Security, LLC for the U.S. DOE's NNSA, Contract #89233218CNA000001. We made use of the Discrete Fourier Transform program FWPEAKS kindly supplied by A. Pigulski, written by Z. Kołaczkowski, G. Kopacki, and W. Hebisch (Astronomical Institute, Wrocław University, Poland).